\documentclass[conference]{IEEEtran}
\IEEEoverridecommandlockouts
\usepackage{cite}
\usepackage{amsmath,amssymb,amsfonts}
\usepackage{graphicx,subfigure}
\usepackage{textcomp}
\usepackage{xcolor}
\usepackage{graphicx}
\usepackage{textcomp}
\usepackage{multicol}
\usepackage{floatrow}
\DeclareMathOperator*{\argmax}{argmax}
\usepackage{algpseudocode}
\usepackage{algorithm}
\usepackage{colortbl}
\onecolumn

\def\BibTeX{{\rm B\kern-.05em{\sc i\kern-.025em b}\kern-.08em
    T\kern-.1667em\lower.7ex\hbox{E}\kern-.125emX}}
\begin{document}

\title{Multi-Step Traffic Prediction for Multi-Period Planning in Optical Networks}

\author{\IEEEauthorblockN{Hafsa Maryam, Tania Panayiotou, Georgios Ellinas}\\
\IEEEauthorblockA{\textit{KIOS CoE and Department of Electrical and Computer Engineering}, \\ \textit{University of Cyprus, Nicosia, Cyprus},\\
\{maryam.hafsa, panayiotou.tania, gellinas\}@ucy.ac.cy}}

\maketitle

\begin{abstract}
\noindent A multi-period planning framework is proposed that exploits multi-step ahead traffic predictions to address service overprovisioning and improve adaptability to traffic changes, while ensuring the necessary quality-of-service (QoS) levels. An encoder-decoder deep learning model is initially leveraged for multi-step ahead prediction by analyzing real-traffic traces. This information is then exploited by multi-period planning heuristics to efficiently utilize available network resources while minimizing undesired service disruptions (caused due to lightpath re-allocations), with these heuristics outperforming a single-step ahead prediction approach.\\
\\
{\bf Keywords:} Traffic prediction; Multi-period planning; Service disruptions; Machine learning; Encoder-Decoders; 
\end{abstract}

\section{Introduction}
Network capacity demand is rapidly increasing, due to the emergence of new services and applications. To cope with this growing demand, the use of machine learning (ML) techniques for traffic-driven service provisioning has emerged as a promising solution to effectively model real-world traffic traces \cite{9748600}  and deal with overprovisioning that is present in statically-provisioned elastic optical networks (EONs) \cite{SurveyArxiv22}. In essence, traffic modeling and prediction enables multi-period planning, as traffic predictions can be used for in-advance network re-optimization to ensure the efficient utilization of available network resources (i.e., avoid overprovisioning), while at the same time respecting the targeted QoS requirements of the end-users (e.g., avoid underprovisioning)~\cite{8620207}. Network traffic modeling involves time-series forecasting by leveraging past traffic traces to predict future traffic demand, subsequently allowing for proactive network optimization (i.e., predictive service provisioning). 

Among the various ML-based models applied for traffic prediction, deep neural networks (DNNs)~\cite{8436062} with recurrent units (i.e., gated recurrent units (GRU), long short-term memory (LSTM) units)~\cite{Balanici:21,9782838} have demonstrated competitive performance accuracy~\cite{SurveyArxiv22}, mainly due to their ability to efficiently capture the nonlinear nature and long-range dependencies of network traffic. For predictive service provisioning, several optimal and heuristic-based algorithms have been developed to exploit the traffic predictions towards reducing service overprovisioning~\cite{9748600, maryam2023uncertainty}, service disruptions~\cite{6831425}, and energy consumption~\cite{Alvizu2017EnergyED}.  

The vast majority of existing work, however, addresses traffic-driven service provisioning by exploiting single-step ahead predictions, with the exploitation of multi-step ahead predictions (e.g., for several hours ahead) remaining greatly unexplored. While the multi-step ahead prediction problem is examined in~\cite{Balanici:21, 9782838}, only in~\cite{9782838} multi-step ahead predictions are used for service provisioning in optical networks. Specifically,~\cite{9782838} uses an encoder-decoder LSTM (ED-LSTM) model to predict future traffic demand matrices  several steps ahead in an optical datacenter and a high-performance computing network. These predictions are subsequently clustered according to an unsupervised learning technique to identify significant variations in future network traffic, ultimately triggering a network reconfiguration (i.e., when a predicted traffic demand matrix is clustered differently from the previous one). That work demonstrates that, leveraging knowledge about traffic variations over multiple future time steps aids in reducing service disruptions and in turn improves end-to-end packet latency and packet loss rates when compared to a static network scenario. 

The focus of this work is to propose and evaluate traffic-driven routing and spectrum allocation (RSA) algorithms by also exploiting multi-step ahead predictions. An ED-LSTM model, designed for sequence-to-sequence problems \cite{cho-etal-2014-learning}, is initially leveraged for multi-step ahead predictions. These predictions are then exploited by the proposed multi-period planning heuristics to minimize undesired service disruptions (due to lightpath re-allocations), while also ensuring that spectrum resources are efficiently utilized. Comparisons with a heuristic exploiting single-step ahead predictions demonstrate that the multi-step approach performs significantly better in terms of service disruptions, with a negligible impact on overprovisioning.  

\begin{figure*}[t!]
\centering
{\includegraphics[scale =0.4]{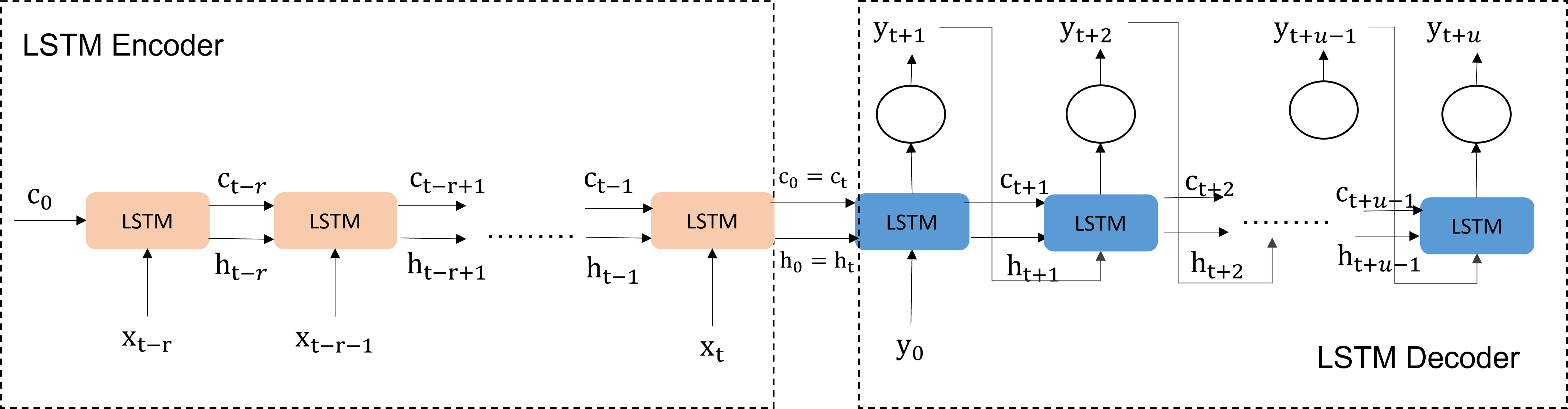} 
\vspace{-0.1in}
\caption{A generic ED-LSTM architecture.}
\label{lstm}}
\end{figure*}

\section{Multi-Step Ahead Traffic Prediction} \label{overview}
The objective of the ED-LSTM model \cite{cho-etal-2014-learning} can be described as learning a non-linear function $f(\cdot)$ to accurately forecast future traffic demands (e.g., bit-rates) based on past and present traffic observations. In this work, past and present traffic observations are given as vector $\chi_t=[x_{t-r}, \cdots, x_{t-1},x_t]$ and future traffic demands are given as vector $\psi_t=[y_{t+1}, y_{t+2}, \cdots, y_{t+u}]$; $r$ is the number of past observations, $t$ is the present time interval during which the current traffic demand $x_t$ is observed, and $u$ is the number of future traffic predictions (i.e., the number of future planning intervals). 

In a generic ED-LSTM architecture (Fig.~\ref{lstm}), the LSTM cell takes an input $x_{t'}$ (i.e., observation of traffic demand at time interval $t'$) and updates its hidden state $h_{t'}$  based on the input and the previous hidden state. In general, the LSTM encoder reads each input $\chi_t$ sequentially to update the cell and hidden states accordingly. After reading the end of each input sequence, the encoder summarizes this input sequence in the cell state and hidden state vectors $c_t$ and $h_t$, respectively, that are then fed as input to the decoder component. The decoder recursively predicts the traffic demand $y_{t+1},.., y_{t+u}$ according to $y_{t'}=g(h_{t'})$ $ \forall t'=t+1,...,t+u$, where $g(\cdot)$ is the activation function of the output layer. In the traffic prediction problem considered in this work, the ED-LSTM is trained as a regressor, thus the decoder's output is followed by a fully connected dense layer.

For training the ED-LSTM model, the Adam optimization algorithm is utilized. Specifically, the model is optimized according to a dataset $D={\{\chi_t, \psi_t\}_{t=1}^{n}}$, consisting of real traffic demand traces (described in Sec.~\ref{dataset}), to reduce the mean squared error (MSE) loss function, where $n$ denotes the observed traffic sequences. After training, the ED-LSTM model predicts future traffic values as $y_{t+u'} = f(x_{t-r},...., x_{t-1},x_{t},y_{t+1},y_{t+2},....,y_{t+u'-1})$, where $u'$ represents any future time interval and $u'\leq u$. Note that, for illustration purposes, the LSTM cells in Fig.~\ref{lstm} unfold in time. However, in practice, the encoder-LSTM cell processes $r$ traffic demand observations in sequence, while the decoder-LSTM cell sequentially predicts future traffic demands $u$. More information about ED models can be found in~\cite{cho-etal-2014-learning}, with our implementation based on the openly available code in~\cite{code}.

\vspace{-0.05in}

\subsection{Dataset Pre-processing} \label{dataset}
The training dataset is created according to the real traffic traces of the Abilene backbone network consisting of $12$ nodes and $30$ links (http://sndlib.zib.de/home.action). This dataset contains bit-rate information (in Gbps) for every node pair in the form of traffic demand matrices, given every $5$-minutes for a $6$-month period. In this work, a dataset $D_v$ is created for each node $v$ in the network (representing the aggregated bit-rate to node $v$ in $5$-minute intervals) to train a different prediction model for each source-destination pair $v-v'$ (to avoid degradation of the model's achievable accuracy when dealing with diverse sequence-to-sequence problems (i.e., with diverse time series)~\cite{10252085}). Given the aggregated bit rates, input traffic patterns for each $D_v$ are created as $\chi_t=[x_{t-r},..., x_{t}]$, where $x_{t-i}=[f_{t-i}^1,...,f_{t-i}^k]$ $\forall i=0,1,...,r$, $\forall t=1,2,...,T$, $t$ is the present planning interval, $r$ is the number of  past planning intervals considered, and vector $x_t$ consists of $k$ traffic fluctuations in $t$. Regarding the ground-truths of each $D_v$, these are created as $\psi_t=[y_{t+1},..., y_{t+u}]$, where $y_{t+j}= \max\{f_{t+j}^1,...,f_{t+j}^k\}, \quad \forall j=0,1,..,u$, $\forall t=1,2..,T$, targeting to predict the maximum traffic demand in $t$ (i.e., to avoid underprovisioning). 

Since in the original dataset bit-rate information is provided in $5$-minute intervals for the creation of the $D_v$ datasets, it is assumed that fluctuations occur on the same time scale. Specifically, it is assumed that a planning interval spans a period of $\tau=30$ minutes, resulting in $k=6$ fluctuations within each planning interval. For input patterns, fluctuations of the previous $r=3$ planning intervals are considered, while $u=4$ to accurately predict traffic demand for the next four planning intervals. For each $D_v$, each sequence $[\chi_t,\psi_t]$ is created following the sliding window approach. 

\begin{figure*}[h!]
 \begin{center}
\includegraphics[clip,scale =0.4]{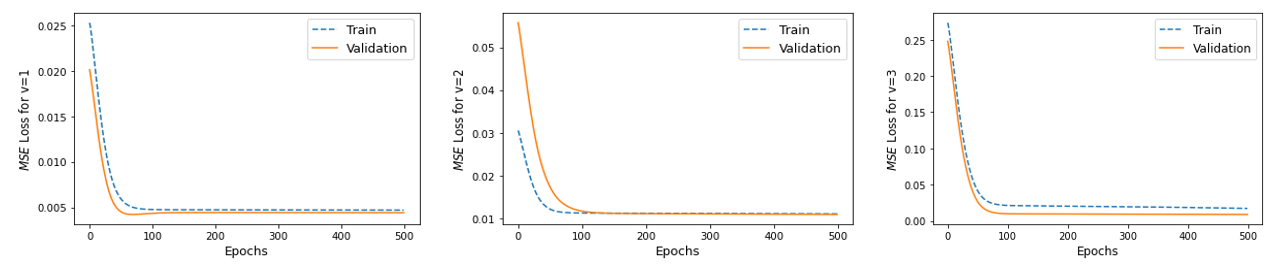}
\vspace{-0.2in}
 \caption{ED-LSTM training evolution vs. MSE loss for $D_1$, $D_2$, and $D_3$ Abilene nodes.}
\label{model}
\end{center}
\end{figure*} 

\vspace{-0.1in}

\subsection{Model Training and Evaluation}
For model training and evaluation, $T$=$800$ sequential traffic patterns are considered within each dataset $D_v$ (spanning the dataset's first $17$ days). All datasets are normalized in the range $[0,1]$. Eighty percent ($80\%$) of the patterns in $D_v$ (the first $640$ patterns) are used for training and validation and $20\%$ ($160$ patterns, spanning $3.5$ days ahead) are used for testing model accuracy. The ED components of the ED-LSTM model are designed according to one hidden layer with $30$ hidden units. The output of the LSTM decoder is wrapped by a time-distributed dense layer with $20$ units. For model training, a learning rate of $0.001$, a batch size of $16$, and $500$ epochs are considered. Note that the ED-LSTM model is tested according to various hyperparameters (e.g., batch size, learning rate) and the configurations chosen are the ones producing the most accurate models.

Figure \ref{model} illustrates the evolution of the ED-LSTM model training using the datasets from source nodes $v=1,2,3$ (i.e., nodes ATLAM5, ATLAng, and CHINng, respectively), demonstrating that the loss function for all models is sufficiently optimized (converging to a sufficiently low test accuracy on their respective test datasets ($0.004$, $0.010$, $0.008$ for $D_1$, $D_2$, and $D_3$, respectively)), with a similar training behavior observed for all $D_{v}$ datasets of the network. Further, it is worth noting that values of $u > 4$ were also examined, with the results indicating that the accuracy of the model starts degrading (since the capabilities of ED-LSTM models to predict over long future horizons are generally limited by the length of the future horizon).

This work achieves to obtain models capable of accurately predicting the traffic demand for $4$ planning intervals ahead of time (i.e., for the next $2$ hours). This information is then exploited by appropriately designed heuristics to reduce undesired service disruptions, while efficiently utilizing the spectrum resources.

\begin{figure}[t]
 \begin{center}
\includegraphics[clip,scale =0.4]{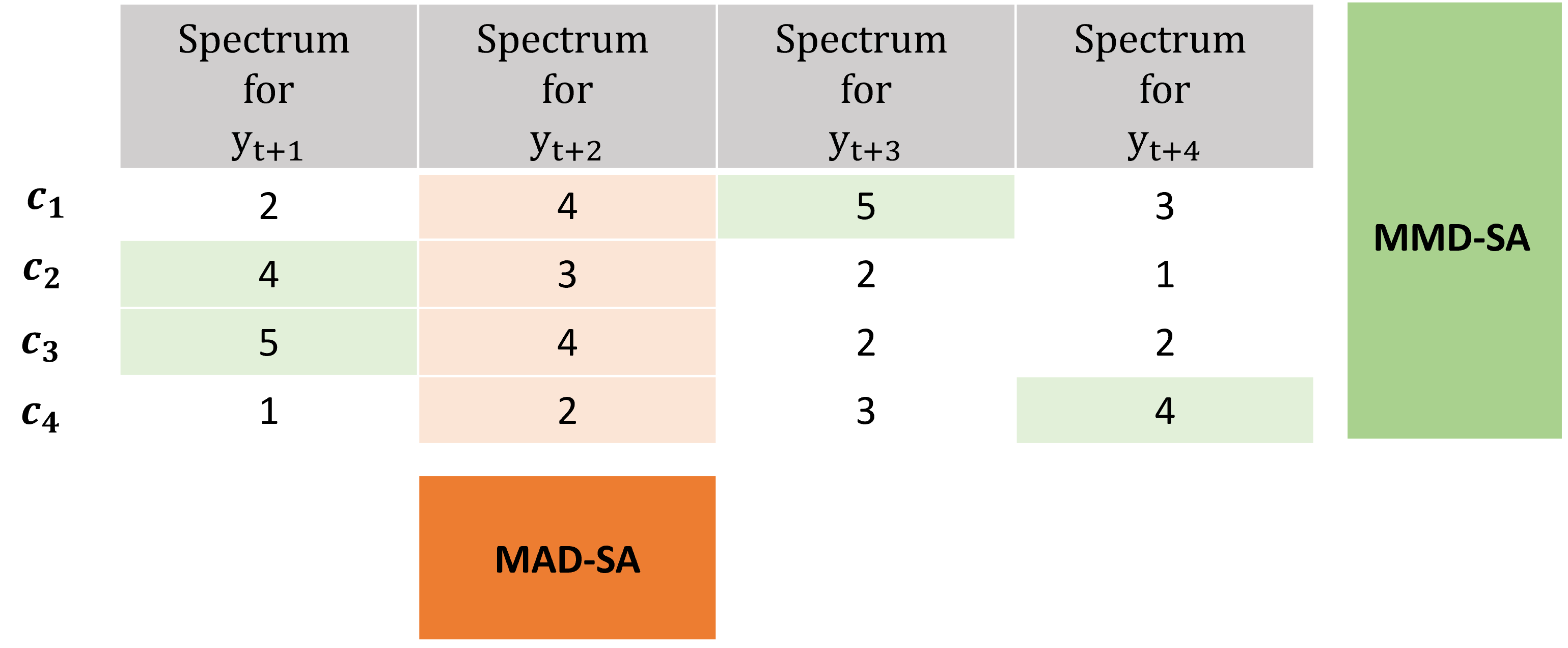}
\vspace{-0.2in}
 \caption{MMD-SA and MAD-SA examples, given the spectrum predicted for each time interval $t+i$ and for each connection $c_i$, with $u=4$.}
 \label{alg_example}
\end{center}
\end{figure} 

\section{Multi-Period Routing and Spectrum Allocation}
To assess the effectiveness of the multi-step ahead prediction approach, different provisioning schemes (i.e., different SA approaches) are examined for an EON. Specifically, two heuristic algorithms are proposed, the multi-step maximum demand spectrum allocation (MMD-SA) and the multi-step average demand spectrum allocation (MAD-SA) heuristics, with each heuristic differently exploiting the multi-step traffic prediction information to select the spectrum size to be allocated for the next $u$ planning intervals. Both heuristics are evaluated and compared against a baseline heuristic, i.e., the single step-ahead demand spectrum allocation heuristic (SSD-SA). Note that SA schemes utilizing single-step predictions are shown to outperform static network provisioning in terms of spectrum savings~\cite{SurveyArxiv22}, hence such a comparison is omitted in this work. The focus is instead on multi-period planning schemes, where fluctuation-driven lightpath re-allocations may cause network disruptions. Thus, the objective is to reduce service disruptions, while efficiently utilizing the available network resources. 

For all heuristics the $\kappa$-shortest path algorithm is utilized to compute the routes ($\kappa=3$), and the shortest path that satisfies the RSA constraints is selected as a feasible lightpath. If no feasible lightpath exists, the connection is blocked. The spectrum (in frequency slots - FSs) for each prediction (in Gbps) is determined using a conventional distance-adaptive modulation scheme based on the predicted bit rates and the distance of the longest path amongst the $\kappa$-shortest paths (taking into consideration that the EON operates with BPSK, QPSK, 8-QAM, and 16-QAM modulation formats \cite{5534599}). Note that for simplicity (i.e., to keep the complexity of MAD-SA low), the distance of the longest path is utilized for all three heuristics (i.e., a worst-case assumption). Then, each heuristic aims to select an appropriate spectrum allocation for the next planning interval(s). Specifically:

\begin{itemize}
\item {\bf  MMD-SA} allocates to each connection request the maximum spectrum computed across all $u$ future planning intervals. As an example, Fig.~\ref{alg_example} indicates the SA decision for each connection according to MMD-SA (i.e., 5,4,5,4 for $c_1,c_2,c_3,c_4$ connections, respectively.). 
\item {\bf MAD-SA} allocates to each connection request the spectrum computed for the $t+i$ future planning interval, where $t+i, i\in {1,..,u}$ is the planning interval for which the aggregated predicted spectrum among all connections is the maximum (i.e., $\argmax_i =\max \{\sum_v y^v_{t+i}\}_{i=1}^u$, where $y^v_{t+i}$ is the spectrum predicted at $t$ for the connection with source node $v$ and for planning interval $t+i$). As an example, Fig.~\ref{alg_example} indicates the SA decision for each connection according to MAD-SA (i.e., 4,3,4,2 for $c_1,c_2,c_3,c_4$, respectively.). Specifically, in this example, MAD indicates that $t+2$ is the planning interval for which the aggregated predicted spectrum is the maximum (i.e., 13 spectrum slots), therefore for each connection the SA predictions for $t+2$ are selected for spectrum allocation. 
\item {\bf SSD-SA} (baseline) assumes single-step ahead predictions (i.e., $u=1$). This heuristic allocates to each connection request the spectrum computed for the next planning interval.
\end{itemize}

For all heuristics, to reduce undesired disruptions due to connection re-allocations driven by traffic fluctuations, spectrum expansion/reduction mechanisms, enabled by the tunable bandwidth variable transponders in EONs~\cite{6381740}, take place whenever possible. Specifically, spectrum reduction occurs when the allocated spectrum is more than the one requested upon a traffic fluctuation (which is always possible) and spectrum expansion is attempted when the already allocated spectrum is less than the one requested upon a traffic fluctuation. If the latter is not possible, connection re-allocation is attempted causing a disruption. Connection blocking occurs if the re-allocation is not achieved. .

\section{Performance Evaluation}\label{evaluation}
The evaluation and comparison of the service provisioning schemes is performed on the Abilene network. The capacity of each fiber link in that network is $320$ FSs, with a spacing of $12.5$ GHz, and a baud rate of $10.5$ Gbaud. Twelve ($12$) ED-LSTM models are used, with each model representing a $v$-$v'$ source-destination pair (destination node $v'$ is randomly selected for each source node $v$). The bit rates for both the predicted and true values are multiplied by $50$ for all schemes to ensure reasonable spectrum demands given the capacity of the network. All schemes are assessed in the test datasets $D_v$, against the true fluctuations (i.e., each SA decision taken according to either MMD-SA, MAD-SA, or SSD-SA for an interval $t$ is compared against the true fluctuations in $x_{t}=[f_{t}^1,...,f_{t}^k]$, with each fluctuation converted to spectrum slots). 

The outcomes of the service provisioning schemes are presented in Table~\ref{results}, which includes information on the number of blocked connections, number of disruptions averaged over the number of connections, and unutilized FSs (overprovisioning) averaged over all fluctuations and across all source-destination pairs. Note that disruptions are calculated according to the total number of times a re-allocation is performed (i.e., a different central frequency is allocated, causing a disruption). 

Based on the results in Table ~\ref{results}, all three heuristics result in zero blocking. Further, both proposed heuristics, MAD-SA and MMD-SA, greatly outperform the baseline SSD-SA approach in terms of service disruptions (i.e., up to $34\%$ ); an indicator of the importance of performing service provisioning by having insights on how the traffic demand varies over a long future horizon (i.e., by considering multi-step ahead predictions). Comparatively, MMD-SA outperforms MAD-SA (by approximately $15\%$) in terms of service disruptions, as in MMD-SA the spectrum allocation considers the maximum predicted traffic, reducing the likelihood of a future traffic demand exceeding the already allocated spectrum. The price of reducing service disruptions is, however, service overprovisioning. As expected, SSD-SA results in fewer (on average) unutilized slots, as in this scheme the connections are subject to more frequent re-allocations in order to more closely follow the traffic fluctuations. On the contrary, both multi-step ahead based schemes increase the average number of unutilized slots (up to $33\%$) for the sake of decreasing service disruptions (up to $34\%$). Between the two proposed multi-step ahead schemes, MAD-SA better balances this trade-off, as it increases overprovisioning by $13\%$ for an improvement of $17\%$ in service disruptions compared to the baseline approach. Overall, the selection between the proposed schemes greatly depends on the end-user's service level agreements (i.e., as it concerns the tolerance on service disruptions that may cause loss of traffic).   

\begin{table}
\centering
\caption{Evaluation of service provisioning schemes.}
\label{results}
\begin{tabular}{|c|c|c|c|}
\hline
& MAD-SA  &  MMD-SA   &  SSD-SA \\ \hline
{Blocked Connections} & 0 & 0 & 0\\ \hline
{Disruptions} & 74.5 & 59.3 & 89.8 \\ \hline
{Unutilized FSs} & 5.35 & 6.97 & 4.64\\ \hline
\end{tabular}
\end{table}

\section{Conclusion}\label{con}
This work investigates the significance of employing an ML-aided technique based on an ED-LSTM model to predict network traffic demand over a long future horizon. Specifically, it is shown that MAD-SA and MMD-SA heuristics, based on multi-step ahead predictions, greatly outperform SSD-SA (based on single-step ahead predictions) in services disruptions (up to $34\%$). Additionally, it is shown that MMD-SA outperforms MAD-SA in terms of service disruptions, albeit with less spectrum savings ($23\%$). Notably, both the MAD-SA and MMD-SA schemes enable network operators to effectively identify and handle the future traffic demand over a longer horizon (e.g., $2$ hours prior), while the selection between the two schemes depends on the end-user's tolerance on service disruptions.  

\section*{Acknowledgements}
This work has been supported by the European Union’s Horizon 2020 research and innovation programme under grant agreement No. 739551 (KIOS CoE - TEAMING) and
from the Republic of Cyprus through the Deputy Ministry of Research, Innovation, and Digital Policy.

\bibliographystyle{IEEEtran}
\bibliography{IEEEabrv,biblio}

\begin{thebibliography}{10}
\providecommand{\url}[1]{#1}
\csname url@samestyle\endcsname
\providecommand{\newblock}{\relax}
\providecommand{\bibinfo}[2]{#2}
\providecommand{\BIBentrySTDinterwordspacing}{\spaceskip=0pt\relax}
\providecommand{\BIBentryALTinterwordstretchfactor}{4}
\providecommand{\BIBentryALTinterwordspacing}{\spaceskip=\fontdimen2\font plus
\BIBentryALTinterwordstretchfactor\fontdimen3\font minus
  \fontdimen4\font\relax}
\providecommand{\BIBforeignlanguage}[2]{{%
\expandafter\ifx\csname l@#1\endcsname\relax
\typeout{** WARNING: IEEEtran.bst: No hyphenation pattern has been}%
\typeout{** loaded for the language `#1'. Using the pattern for}%
\typeout{** the default language instead.}%
\else
\language=\csname l@#1\endcsname
\fi
#2}}
\providecommand{\BIBdecl}{\relax}
\BIBdecl

\bibitem{9748600}
T.~Panayiotou and G.~Ellinas, ``Addressing traffic prediction uncertainty in
  multi-period planning optical networks,'' in \emph{Proc. IEEE/OSA Optical
  Fiber Communications Conference (OFC)}, 2022.

\bibitem{SurveyArxiv22}
T.~Panayiotou, M.~Michalopoulou, and G.~Ellinas, ``Survey on machine learning
  for traffic-driven service provisioning in optical networks,'' \emph{IEEE
  Communications Surveys \& Tutorials}, vol.~25, no.~2, pp. 1412--1443, 2023.

\bibitem{8620207}
T.~Panayiotou, K.~Manousakis, S.~Chatzis, and G.~Ellinas, ``A data-driven
  bandwidth allocation framework with {Q}o{S} considerations for {EON}s,''
  \emph{IEEE/OSA J. Light. Technol.}, vol.~37, no.~9, pp. 1853--1864, 2019.

\bibitem{8436062}
X.~Chen, R.~Proietti, H.~Lu, A.~Castro, and S.~Yoo, ``Knowledge-based
  autonomous service provisioning in multi-domain elastic optical networks,''
  \emph{IEEE Commun. Mag.}, vol.~56, no.~8, pp. 152--158, 2018.

\bibitem{Balanici:21}
M.~Balanici and S.~Pachnicke, ``Classification and forecasting of real-time
  server traffic flows employing long short-term memory for hybrid {{E/O}} data
  center networks,'' \emph{IEEE/OSA J. Opt. Commun. Netw.}, vol.~13, no.~5, pp.
  85--93, 2021.

\bibitem{9782838}
S.~K. Singh, C.-Y. Liu, S.~J. Ben~Yoo, and R.~Proietti,
  ``Machine-learning-aided dynamic reconfiguration in optical {DC}/{HPC}
  networks,'' in \emph{Proc. IEEE Optical Network Design and Modelling (ONDM)},
  2022.

\bibitem{maryam2023uncertainty}
H.~Maryam, T.~Panayiotou, and G.~Ellinas, ``Uncertainty quantification and
  consideration in {ML}-aided traffic-driven service provisioning,''
  \emph{Comp. Commun.}, vol. 202, pp. 13--22, 2023.

\bibitem{6831425}
S.~Shakya, Y.~Wang, X.~Cao, Z.~Ye, and C.~Qiao, ``Minimize sub-carrier
  reallocation in elastic optical path networks using traffic prediction,'' in
  \emph{Proc. IEEE Global Communications Conference (GLOBECOM)}, 2013.

\bibitem{Alvizu2017EnergyED}
R.~Alvizu, X.~Zhao, G.~Maier, Y.~Xu, and A.~Pattavina, ``Energy efficient
  dynamic optical routing for mobile metro-core networks under tidal traffic
  patterns,'' \emph{IEEE/OSA J. Light. Technol.}, vol.~35, no.~2, pp. 325--333,
  2017.

\bibitem{cho-etal-2014-learning}
K.~Cho, B.~Van~Merri{\"e}nboer, C.~Gulcehre, D.~Bahdanau, F.~Bougares,
  H.~Schwenk, and Y.~Bengio, ``Learning phrase representations using {RNN}
  encoder-decoder for statistical machine translation,'' \emph{arXiv:1406.1078
  [cs.CL]}, 2014.

\bibitem{code}
\BIBentryALTinterwordspacing
J.~Yoon. Time-series prediction with { RNN}, {GRU}, {LSTM} and attention.
  [Online]. Available:
  \url{https://github.com/jsyoon0823/Time-series-prediction}
\BIBentrySTDinterwordspacing

\bibitem{10252085}
S.~Behera, T.~Panayiotou, and G.~Ellinas, ``Machine learning framework for
  timely soft-failure detection and localization in elastic optical networks,''
  \emph{IEEE/OPTICA J. Opt. Commun. Netw.}, vol.~15, no.~10, pp. E74--E85,
  2023.

\bibitem{5534599}
M.~Jinno, B.~Kozicki, H.~Takara, A.~Watanabe, Y.~Sone, T.~Tanaka, and
  A.~Hirano, ``Distance-adaptive spectrum resource allocation in
  spectrum-sliced elastic optical path network,'' \emph{IEEE Commun. Mag.},
  vol.~48, no.~8, pp. 138--145, 2010.

\bibitem{6381740}
K.~Christodoulopoulos, I.~Tomkos, and E.~Varvarigos, ``Time-varying spectrum
  allocation policies and blocking analysis in flexible optical networks,''
  \emph{IEEE J. Sel. Areas Commun.}, vol.~31, no.~1, pp. 13--25, 2013.

\end{thebibliography}
\end{document}